\documentclass[10pt,aps,twocolumn,groupedaddress,showpacs]{revtex4}
\usepackage{graphicx,amsmath,amsfonts}

\begin{document}

\title{Electron-Electron Relaxation Effect on Auger Recombination in
Direct Band Semiconductors}

\author{Anatoli Polkovnikov}
\email{anatoli.polkovnikov@yale.edu}
\homepage{http://pantheon.yale.edu/~asp28}
\altaffiliation{also at Ioffe Physico-Technical Institute}
\affiliation{Physics Department  Yale University}

\author{Georgy Zegrya}
\email{zegrya@theory.ioffe.rssi.ru}
\homepage{http://www.ioffe.rssi.ru/Dep_TM/zegrya.html}
\affiliation{Ioffe Physico-Technical Institute}

\begin{abstract}
Influence of electron-electron relaxation processes on Auger
recombination rate in direct band semiconductors is investigated.
Comparison between carrier-carrier and carrier-phonon relaxation
processes is provided. It is shown that relaxation processes are
essential if the free path length of carriers doesn't exceed
a certain critical value, which exponentially increases with
temperature. For illustration of obtained results a typical InGaAsP
compound is used.
\end{abstract}


\maketitle

It is well known that the intraband relaxation of carriers plays an
important role in recombination processes~\cite{7,4}. In
particular, relaxation was shown to cause broadening of gain and
emission spectra in semiconductor lasers~\cite{7}. Influence of
relaxation processes on Auger recombination (AR) in bulk
semiconductors is more fundamental. Both for the CHCC and for
the CHHS processes AR coefficient calculated in the first order of
perturbation theory in electron-electron interaction is of the
threshold type, being an exponential function of temperature~\cite{1,2}.
Relaxation eliminates the threshold conditions enhancing AR. Phonon
and impurity assisted Auger processes in A$_{\rm III}$B$_{\rm V}$ homogeneous
semiconductors were studied in ~\cite{Tak1,Tak2,Yevick}. As
shown in~\cite{7}, it is the electron (hole)-electron (hole)
relaxation mechanism which gives the main contribution to
broadening of the light emission spectra in semiconductors with
high carrier density. However, influence of this relaxation
mechanism on AR has not been studied yet. Since AR dominates over
other recombination processes at high
densities of electrons and holes and the role of the carrier-carrier
scattering also increases with carrier concentration, the corresponding
mechanism of relaxation might be expected to be of primary
importance in calculating AR rate.

In this paper we study temperature and carrier-density
dependences of Auger coefficient with and without regard to
electron (hole)-electron (hole) relaxation processes. General
Green function approach is used for calculation of AR coefficient.
Wave functions and energy spectra of electrons and holes are
found from the conventional $8\times8$ $\bf kp$-model.
Comparison of the direct AR, phonon-assisted AR, and AR with
carrier-carrier relaxation is provided both for the CHCC and CHHS
processes.

Finite temperature Green function techniques for calculating
AR rate was developed in~\cite{Tak1,Tak2}.
The formalism used in those papers is based on linear response
and mean-field approximations. It was shown that the relaxation
processes eliminate the threshold appearing in the first order of
perturbation theory on Coulomb interaction and enhance AR rate.
In~\cite{Tak1,Tak2}, there were studied Auger processes
with relaxation on phonons and impurities. However, wave
functions and overlap integrals were phenomenologically
assumed than calculated, which drastically affected the results.
The most accurate calculation of the phonon-assisted AR was performed
in~\cite{Yevick}. But the heavy hole wave functions were not
derived there consistently with those of electrons.

In direct band A$_{\rm III}$B$_{\rm V}$ semiconductors the electron
effective mass is usually much less then that of heavy
holes~\cite{4}. This allows us to neglect by electron scattering
processes, compared with those of holes~\cite{7}, and to use
free particle propagators for the former. By the same reason
we also neglect by the momenta and energies of the electrons in
the initial state~\cite{2}.

Following the approach developed in~\cite{Tak1,Tak2} and using
wave functions derived in $8\times 8$ model~\cite{prb} we obtain
Auger coefficient for the CHCC process:
\begin{widetext}
\begin{eqnarray}
C&\approx& {32\sqrt{2\pi^5}e^4\hbar^2\langle E_{c}\rangle\over
9m_h^{3/2}T^{3/2}E_g\epsilon_\infty^2}
{3E_g+2\Delta_{so}\over E_g+\Delta_{so}}
\int\limits_{-\infty}^\infty
{d{\cal E}\over k_c^2({\cal E}+E_g)}{dk_c\over d{\cal E}}
\exp{\left(-{{\cal E}\over T}\right)}D(k_c({\cal E}+E_g),{\cal E})
\label{chcc}
\end{eqnarray}
\end{widetext}
Here $\langle E_{c}\rangle$ is the mean electron energy equal
to ${3\over 2} T$ if they have Boltzmann distribution, $m_h$ is
the heavy-hole effective mass, $T$ is the temperature in energy
units, $E_g$ and $\Delta_{so}$ are the band gap and spin-orbital
splitting respectively, $k_c(E)$ is the wave vector versus
energy in the conduction band, $E_h(k)$ is the dispersion
of the heavy holes, $D(k,E)$ is the spectral function related to
the imaginary part of the heavy-hole proper self-energy $\Gamma(k,E)$
by
\begin{equation}
D(k,E)={1\over \pi}{\Gamma(k,E)\over
\left(\Gamma(k,E)\right)^2+(E-E_h(k))^2}.
\label{101}
\end{equation}
In (\ref{chcc}) we used Boltzmann distribution of heavy holes, which
is usually the case due to large value of their effective mass.
Generalization of this expression to the case of Fermi statistics is
straightforward. The high frequency dielectric
constant $\epsilon_\infty$ is taken away from the integrand because
the free-carrier screening effects are weak and unimportant for
Auger process~\cite{3c}. We neglected by the real part of the
proper self energy in (\ref{101}),  since its only effect is the
slight renormalization of the heavy hole mass.
It should be noted, that $\Gamma$ strongly damps in the band gap
${\cal E}<0$ and the main contribution to the integral comes from
the positive values of the hole energy.

Neglecting by relaxation processes leads to a well known
expression for the Auger coefficient~\cite{2,3c}: %
\begin{eqnarray}
C\!\approx\! {8\sqrt{2\pi^5}e^4\hbar^3\langle E_{c}\rangle \over
3m_h^{3/2}m_c^{1/2}T^{3/2}E_g^{5/2}\epsilon_\infty^2}\exp{\left(-
{E_{th}\over T}\right)}F\left({\Delta_{so}\over E_g}\right),\phantom{XX}
\end{eqnarray}
where $F(x)$ is a multiplier of the order of $1$~\cite{2}:
\begin{displaymath}
F(\alpha)=\left({1+\alpha\over 1+2\alpha/3}\right)^{3/2}
\left({1+\alpha/3\over 1+\alpha/2}\right)^{1/2},
\end{displaymath}
$E_{th}\approx 2m_c/m_h E_g$ is the threshold energy.

Similarly can be obtained the expression for the CHHS Auger process:
\begin{eqnarray}
&&C={16\pi^2e^4\hbar^5\over 3\epsilon_\infty^2 m_h^3 T^3}{\langle
E_c\rangle \over E_g}{3E_g\!+\!2\Delta_{so}\over 3(E_g\!+\!\Delta_{so})}
\int\limits_0^\infty \int\limits_0^\infty dk_1dk_2 \int\limits_{-1}^1
d\cos{\vartheta}\times\nonumber\\
&&{(1+\lambda_{so})^2\over 1+2\lambda_{so}^2\!+{\Delta_{so}\over
E_g-E_{so}\phantom{X^X}\hspace{-0.5cm}}{2\lambda_{so}^2\!+\lambda_{so}\!
-1\over 3\lambda_{so}}}
{\cos^2(\vartheta)\,k_1^2\over ({\bf k}_1
+{\bf k}_2)^2}\left[1\!+\!{k_2\over 2k_1}\sin{\vartheta}
\right]\times\nonumber\\
&&\int\limits_{-\infty}^{\infty} dE
D(k_1,E)D(k_2,E_{so}\!-\!E_g\!-\!E)
\exp{\left({Eg-E_{so}\over T}\right)}\label{f},
\end{eqnarray}
where $E_{so}\equiv E_{so}(|{\bf k}_1+{\bf k}_2|)$ is the energy of the
split-off hole, $\vartheta$ is the angle between the wavevectors
${\bf k}_1$ and ${\bf k}_2$,
\begin{displaymath}
\lambda_{so}={\Delta_{so}\over 3\left(E_{so}+4/3\,\Delta_{so}
+{\hbar^2 k_{so}^2(E_{so})\over 2m_h}\right)}.
\end{displaymath}

The presence of the multiplier $(1+\lambda_{so})^2$ in the
integrand ensures that near the $\Gamma$-point (at the center of
Brillouin zone), where $E_{so}\approx -\Delta_{so}$ and
$\lambda_{so}\approx -1$, the overlapping between heavy holes
and split-off holes vanishes~\cite{8}. Note that the exchange
interaction term doesn't vanish for the CHHS process as it does
in the case of the CHCC one.
Namely, this interaction is responsible for the second term in the
square brackets in the integrand of (\ref{f}). If the
spectral functions are substituted by $\delta$-functions,
(\ref{f}) turns to the expression derived by Ge'lmont
et.al.~\cite{8}.

Let us now consider the relaxation processes in detail. In this
paper we will study two basic scattering mechanisms in undoped
semiconductors: (i) on polar optical phonons and
(ii) on electron hole plasma. The imaginary part
of the proper self energy for the first mechanism
with account of the complex valence band structure is as follows:
\begin{eqnarray}
&&\Gamma_{ph}(k,E)={m_h e^2 \omega_{lo}\over 4\hbar\overline{\epsilon} k}
\Biggl[{1\over \exp{\hbar\omega_{lo}\over T}-1}\times\nonumber\\
&&\int\limits_{(1-\Delta_1)^2}^{(1+\Delta_1)^2}
{3(\Delta_1^2+1-\xi)^2+4\Delta_1^2
\over 16\Delta_1^2 \xi}d\xi + {\exp{\hbar\omega_{lo}\over T}\over
\exp{\hbar\omega_{lo}\over T}-1}\times\nonumber\\
&&\int\limits_{(1-\Delta_2)^2}^{(1+\Delta_2)^2}
{3(\Delta_2^2+1-\xi)^2+4\Delta_2^2
\over 16 \Delta_2^2 \xi}d\xi
\Biggr],
\label{ph}
\end{eqnarray}
$\omega_{lo}$ is the phonon frequency assumed to be independent
of the wavevector,
\begin{displaymath}
\Delta_{1,2}\equiv\Delta_{1,2}(k,E)=\left\{\begin{array}{cl}
\sqrt{2m_h(E\mp\hbar\omega_{lo})\over \hbar^2k^2} & \mbox{if}\;
E>\pm\hbar\omega_{lo}\\ 0&\mbox{otherwise}\end{array}\right.,
\end{displaymath}
with indices $1,2$ corresponding to upper, lower sign respectively,
\begin{displaymath}
{1\over \overline{\epsilon}}={1\over \epsilon_\infty}-{1\over \epsilon_0},
\end{displaymath}
$\epsilon_0$ is the low-frequency dielectric constant. The first
and the second terms in (\ref{ph}) correspond to phonon
absorption and phonon emission, respectively. If screening
effects are taken into account than in both terms of (\ref{ph})
$\xi^{-1}$ should be substituted by $\xi/(\xi+\lambda_{TF}^2/k^2)^2$
with $\lambda_{TF}$ being Thomas-Fermi screening momentum. Integration
in (\ref{ph}) can be conducted explicitly however the resulting expression
is quite cumbersome. A similar to (\ref{ph}) formula was derived in
~\cite{Yevick}, but there is some discrepancy due to more accurate
consideration of the heavy-hole spectrum in this paper.

Let us consider the relaxation due to hole scattering on
equilibrium electron-hole plasma. Using RPA approach for calculating
the imaginary part of the self energy $\Gamma_e$ due to
carrier-carrier scattering~\cite{Fetter} we obtain:
\begin{widetext}
\begin{eqnarray}
\Gamma_{e}(k,E)=&&-{me^2\over \epsilon_0 \pi k}\int\limits_0^\infty
d{\cal E}\int\limits_{q_{min}}^{q_{max}} {dq\over q}
\left({1\over \exp{\left({{\cal E}-\mu_v\over T}\right)}+1}
+{1\over \exp{\left({{\cal E}-E\over T}\right)}-1}\right)\times\nonumber\\
&&{\delta\epsilon^{\prime\prime}(q,E-{\cal E})\over
(1+\delta\epsilon^\prime(q,E-{\cal E}))^2+
\left(\delta\epsilon^{\prime\prime}(q,E-{\cal E})\right)^2}
\left[1-{3\over 4}{(E_h(q)-E_h(q_{min}))(E_h(q_{max})-E)
\over 4E_h(k){\cal E}}\right],
\label{ee}
\end{eqnarray}
\begin{displaymath}
q_{min}=\left| \sqrt{2m_h{\cal E}\over\hbar^2}-k\right|,\qquad
q_{max}=\sqrt{2m_h{\cal E}\over\hbar^2}+k.
\end{displaymath}
Here $\delta\epsilon$ is the contribution to the dielectric
constant from free carriers, double prime refers to the
imaginary part and single prime does to the real part of
$\delta\epsilon$. The last multiplier in (\ref{ee}) comes from
the complex structure of the valence band,
\begin{eqnarray}
\delta\epsilon^{\prime\prime}(k,E)=-{2m_h^2e^2\over\epsilon_0^2k^3}\!\!\!
\int\limits_{(E_h(k)-E)^2\over 4E_h(k)}^{\infty}\!\!\! dE_q
(f_h(E_q)-f_h(E_q+E))\left[1-{3E_h(k)\over 4E_q}
{E_q-{(E_h(k)-E)^2\over 4E_h(k)}\over E_q+E}\right],
\label{eps2}
\end{eqnarray}
\end{widetext}

where
\begin{displaymath}
f_h(E)={1\over \exp\left({E-\mu_h\over T}\right)+1}
\end{displaymath}
is the heavy hole distribution function.
The expression for the real part of $\delta\epsilon$ is quite
complicated~\cite{3c}. However, as numerical computation shows,
Thomas-Fermi approximation gives very reasonable results.
So $\delta\epsilon^{\prime\prime}$ in denominator of (\ref{ee}) can
be set zero and
\begin{equation}
\delta\epsilon^\prime=\lambda_{TF}^2/q^2,
\label{eps1}
\end{equation}
where $\lambda_{TF}$ is the inverse Thomas-Fermi screening
length~\cite{Tak1}:
\begin{equation}
\lambda_{TF}\!=\sqrt{{4\sqrt{2}e^2\sqrt{T}\over \pi\hbar^3\epsilon_0}
\left(m_h^{3/2}S\left({\mu_h\over T}\right)\!+m_c^{3/2}S\left({\mu_c
\over T}\right)\right)},
\end{equation}
where
\begin{displaymath}
S(x)=\int\limits_0^\infty dy{1\over \exp(y^2-x)+1}.
\end{displaymath}

Substituting (\ref{eps2}) and (\ref{eps1}) into (\ref{ee}) gives
the final expression for the lifetime of holes in the case of
carrier-carrier scattering.

\begin{figure}
\label{fig1}
\includegraphics[angle=90, width=8.6cm, height=6.5cm]{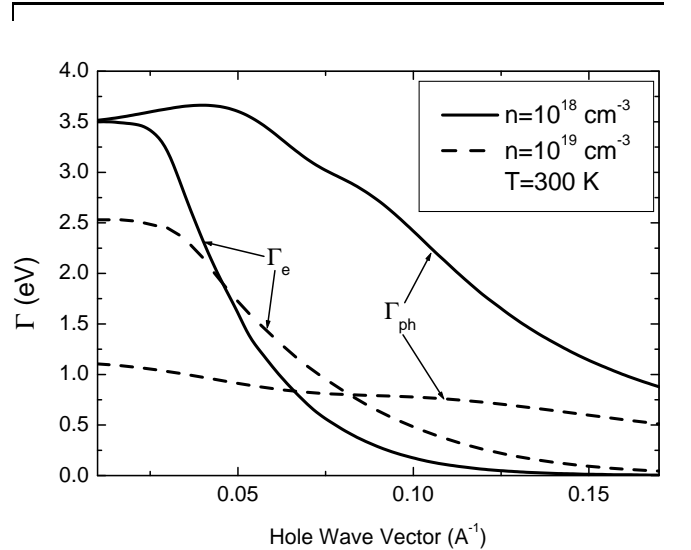}
\caption{Imaginary part of the proper self energy for scattering
of a heavy hole on the electron-hole plasma ($\Gamma_{e}$) and on
the longitudinal polar optical phonos ($\Gamma_{ph}$) as a function of
the heavy hole momentum ($k$) at ($E=0$).}
\end{figure}

\begin{figure}
\label{fig2}
\includegraphics[angle=90, width=8.6cm, height=6.5cm]{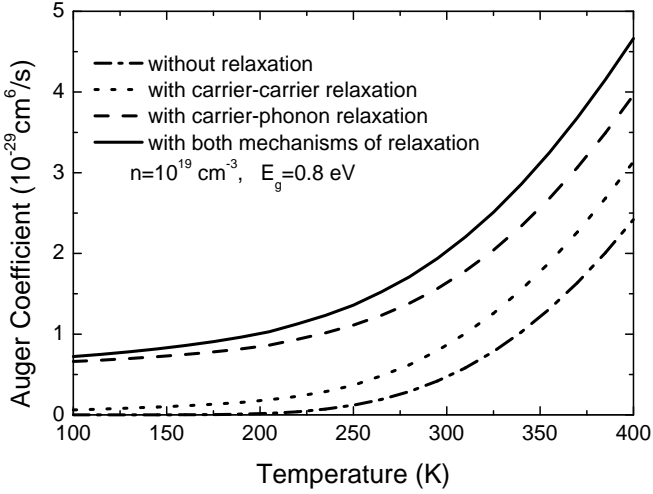}
\caption{CHCC Auger coefficient versus temperature with regard to
different relaxation processes. Parameters of an InGaAsP compound
lattice matched to InP were used in calculations.}
\end{figure}

Imaginary part of the proper self energy $\Gamma_{e}(k,E)$ strongly
depends on both arguments having a steep maximum at $E_h(k)=E$ and
rapidly decreasing when the latter inequality is broken. There are
two main reasons for reducing $\Gamma_e$ when $E_h(k)\neq E$:
(i) Coulomb interaction is relatively weak at large transferred momenta,
(ii) there is an exponentially small number of carriers with large
momenta in equilibrium electron-hole plasma. On the other hand,
scattering on phonons is almost independent of the transferred momentum
and therefore $\Gamma_{ph}$ is a smooth function of its arguments.
Figure 1 shows dependences $\Gamma_e(k)$ and $\Gamma_{ph}(k)$ at a fixed
value of $E=0$. While the value $\Gamma_e$ is larger then that of
$\Gamma_{ph}$ at small values of $k$ and relatively high carrier
densities, the inverse relation could be observed at large $k$ or
small densities. The role of carrier-carrier scattering obviously
increases with carrier density and temperature. In AR, large transferred
momenta play a crucial role~\cite{2,3c}. Therefore the carrier-carrier
scattering mechanism is less important here than in radiative recombination.
Nevertheless this relaxation process remains effective at high
temperatures and carrier densities (Figs. 2-4). Parameters of a
typical InGaAsP compound lattice matched to InP were used for the
illustration of obtained results.

\begin{figure}
\label{fig3}
\includegraphics[angle=90, width=8.5cm, height=6.5cm]{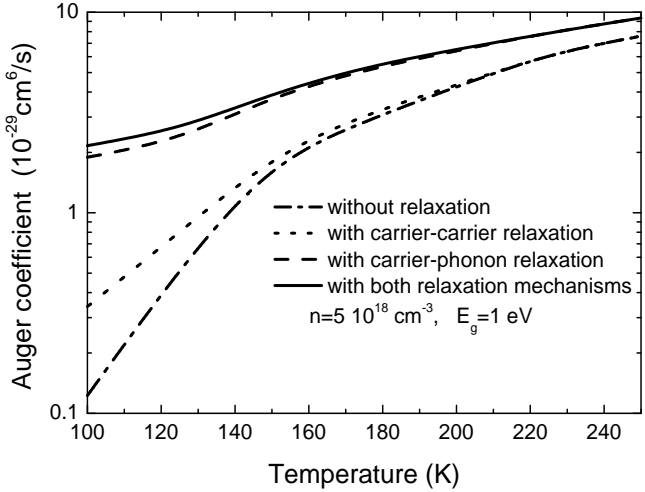}
\caption{CHHS Auger coefficient versus temperature with regard to
different relaxation processes.}
\end{figure}

\begin{figure}
\label{fig4}
\includegraphics[angle=90, width=8.5cm, height=6.5cm]{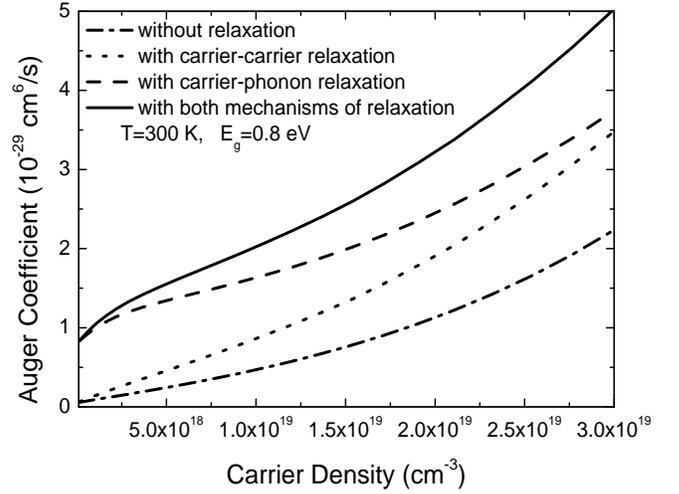}
\caption{ Dependence of CHCC Auger coefficient on carrier density with
and without account of various scattering processes.}
\end{figure}

The threshold energy for the CHCC AR is considerably larger than that
for the CHHS process, therefore relaxation processes are important
for the former up to very high temperature (Fig. 2),
while for the latter they usually give a considerable effect only
when $T$ is small (Fig. 3). At $T$ close to zero phonon-assisted AR
predominates, while the role pf AR with carrier-carrier relaxation
increases at higher $T$ (Fig. 2). The carrier-carrier relaxation mechanism
is also responsible for a stronger dependence of AR coefficient on the
carrier density (Fig. 4).

In conclusion we note that owing to relaxation processes AR becomes
thresholdless, because the restrictions imposed by the energy-momentum
conservation are lifted. From (\ref{chcc}) we can estimate the
characteristic scattering length $a_c$ for the CHCC process
\begin{equation}
a_c=\lambda_{Eg}\left(T\over E_{th}\right)\exp{\left(
E_{th}\over T\right)},
\label{100}
\end{equation}
where $\lambda_{E_g}=2\pi/k_c(E_g)$. If the free path length $\lambda$
exceeds $a_c$ the relaxation is not important, otherwise they should be
taken into account. Note that the similar result was derived in~\cite{prb}
for the case of a quantum well. The role of the momentum relaxation mechanism
there was played by scattering on heterobarriers. The characteristic
quantum well width above which the bulk approximation for the AR is valid
was found to be:
\begin{equation}
\tilde{a}_c=\lambda_{Eg}\left(T\over E_{th}\right)^{3/2}\exp{\left(
E_{th}\over T\right)},
\end{equation}
\newpage
that is almost equal to (\ref{100}). The unessential difference in the
exponent of $T/E_{th}$ is attributed to $1D$ scattering in the case of a
quantum well.

\end{document}